# Persuasion and Phishing: Analysing the Interplay of Persuasion Tactics in Cyber Threats

Kalam Khadka

Faculty of Science and Technology, University of Canberra

***Abstract*—This study extends the research of Ferreira and Teles (2019), who synthesized works by Cialdini (2007), Gragg (2003), and Stajano and Wilson (2011) to propose a unique list of persuasion principles in social engineering. While Ferreira and Teles focused on email subject lines, this research analyzed entire email contents to identify principles of human persuasion in phishing emails. This study also examined the goals and targets of phishing emails, providing a novel contribution to the field. Applying these findings to the ontological model by Mouton et al. (2014) reveals that when social engineers use email for phishing, individuals are the primary targets. The goals are typically unauthorized access, followed by financial gain and service disruption, with Distraction as the most commonly used compliance principle. This research highlights the importance of understanding human persuasion in technology-mediated interactions to develop methods for detecting and preventing phishing emails before they reach users. Despite previous identification of luring elements in phishing emails, empirical findings have been inconsistent. For example, Akbar (2014) found 'authority' and 'scarcity' most common, while Ferreira et al. (2015) identified 'liking' and 'similarity.' In this study, 'Distraction' was most frequently used, followed by 'Deception,' 'Integrity,' and 'Authority.' This paper offers additional insights into phishing email tactics and suggests future solutions should leverage socio-technical principles. Future work will apply this methodology to other social engineering techniques beyond phishing emails, using the ontological model to further inform the research community.**

## I. INTRODUCTION

Phishing is a form of cybercrime that involves impersonating a trustworthy entity to trick users into disclosing sensitive information or performing malicious actions [1]. Phishing involves using a blend of social engineering tactics and technical strategies to persuade individuals to disclose their personal information [2]. Social engineering attacks, particularly phishing, pose significant cybersecurity threats by exploiting human vulnerabilities to obtain sensitive information [2] [3]. Social engineering exploits human vulnerabilities using persuasion principles to gather information and conduct cyber-attacks [4]. Moreover, Phishing emails using social engineering techniques are effective at exploiting human vulnerabilities [5]. Human vulnerabilities pose significant challenges in cybersecurity, with studies indicating that 95% of successful cyber-attacks result from human error [6]. Contextualizing phishing emails to appeal to psychological weaknesses increases susceptibility to deception and vulnerability to phishing [7]. Humans are the weakest link in cybersecurity due to vulnerabilities exploited by social engineering tactics [8]. Integrating human factors with technical solutions can enhance cybersecurity by addressing user vulnerabilities and improving security culture [9]. Phishing, particularly through email delivery, remains a prevalent and costly method of cyberattack. According to ACCC (2024) report, phishing scams have resulted in financial losses amounting to $5,437,833 in Australia from Jan to May 2024, the total amount lost to various scams including phishing, reached $113,270,400, reflecting the substantial economic impact of these cyber threats. Despite comprising only 7% of all scam reports, these phishing attempts have demonstrated a high effectiveness in extracting financial information from victims [10]. The persistent use of phishing emails demonstrates their potency in deceiving individuals by mimicking legitimate communications to harvest sensitive information and execute financial fraud.

Phishing emails are successful because they leverage social engineering techniques, exploiting human psychology and social interactions [5] [11]. They appear credible by using personal and contextual details about their targets, effectively persuading victims to disclose sensitive information and transfer money [12]. The author suggest that phishing emails often incorporate various persuasion principles and techniques to enhance their effectiveness. Therefore, analysing and identifying these principles and behavioural traits in emails can potentially improve and support existing phishing detection tools and making aware about the cyber security threats.

Persuasion, commonly associated with marketing and sales, is a powerful tool that can be applied to any human interaction, including cyber-attacks. It is often used to create convincing phishing emails [1] [11]. Cialdini (2007), Gragg (2003), and Stajano and Wilson (2011) have outlined various persuasion principles frequently employed by humans. Ferreira and Teles (2019) have integrated the three perspectives to derive a unique, complete, and systematized list of principles.

Phishing emails employ various persuasion techniques to deceive recipients. Analysing the content and persuasive elements of phishing emails can provide insights into how they are designed to be persuasive [13] [14]. Phishers often use email to launch their attacks, relying on persuasion to trick unsuspecting victims into responding positively. To protect users, it's crucial for system designers and security professionals to understand how these persuasive techniques work in phishing emails. This study analyses these techniques to better understand how phishing emails operate in the real world.

This study builds upon Ferreira & Teles' (2019) foundational research, which synthesized key principles of persuasion from the works of Cialdini (2007), Gragg (2003), and Stajano and Wilson (2011). Using a coding framework derived deductively from these integrated principles, the current research conducted content analysis on a random selection of 200 phishing emails. These emails were sourced from millerssmiles.com.uk, a reputable international repository of reported spoof emails and phishing scams spanning from 2014 to 2023.

## II. Principles of persuasion

Persuasion is the art of effective speaking and writing to influence the audience through logic, emotion, and credibility, it is also a goal-oriented use of language aimed at influencing attitudes and behaviours [15]. The principle of persuasion has been a subject of research for decades, with various studies exploring its mechanisms and applications. Early studies focused on identifying factors influencing message processing in persuasive communications [16]. Information systems researchers have also recognized the importance of persuasion theory in examining how technology influences attitudes and behaviours. To address the complexity of the persuasion literature, efforts have been made to develop a common frame of reference for conceptualizing persuasion and differentiating it from related concepts in IS research [17]. Principles of persuasion are fundamental concepts in social psychology and communication that explain how individuals can influence others' attitudes and behaviours. These principles can be applied to various contexts, including mass persuasion, as explored by [18]. Cialdini (1993) identified six universal principles of influence, which have become widely recognized in the field. These key principles, known as the theory of influence, claimed as encompassing all human persuasion techniques: authority, reciprocity, commitment and consistency, social proof, liking and similarity, and scarcity [19]. In 2003, Gragg examined the psychological triggers that contribute to the success of social engineering, such as: strong affect, overloading, reciprocation, deceptive relationships, diffusion of responsibility and moral duty, authority, and integrity and consistency [20]. In 2009, Stajano and Wilson analysed various scams that were investigated, documented, and recreated for a BBC TV program. They identified general principles based on recurring victim behaviour patterns that scammers exploit. These principles include Social Compliance, Herd, Deception, Dishonesty, Time, Need and Greed, and Distraction [21]. Together, these studies provide a comprehensive understanding of persuasion techniques and their applications in various domains. The principles of persuasion are influential factors in human decision-making, applicable in various contexts including marketing, social engineering, and phishing attacks [22].

### A. *Unique list of principles of persuasion in Social Engineering*

Ferreira and Teles (2019) proposed developing a comprehensive list of persuasion principles to understand how phishing emails influence users and bypass security measures. They created a unique list of persuasion principles in social engineering by analysing the relationships between the works of Cialdini (2007), Gragg (2003), and Stajano and Wilson (2011). The final list includes five principles, collectively named the Principles of Persuasion in Social Engineering.

P1: Authority - Society conditions individuals to obey authority without question [11]. People typically follow an expert or an authoritative figure and will go to great lengths for someone they perceive as being in charge[11]. For example, an email that appears to be from the recipient's bank and includes the bank's name in the subject line exploits this principle.

P2: Social Proof - People often follow the actions of the majority, lowering their guard and suspicion to share the same responsibilities and risks [11]. This way, they avoid sole accountability if something goes wrong [11]. For example, an email from a purported system administrator with a company email address asks the recipient to test a link, claiming that their colleagues are also testing it.

P3: Liking, Similarity, and Deception - People are more inclined to follow or relate to individuals they know, like, find attractive, or perceive as like themselves [11]. However, appearances can be deceiving, and individuals are often manipulated into believing false identities [11]. For example, an email from someone pretending to be a friend of the recipient, asking them to visit an interesting website, exploits this principle.

P4: Distraction - When people are preoccupied with potential gains, losses, urgent needs, intense emotions, or the scarcity of an item, they often overlook other important factors in their decision-making [11]. For instance, an email claiming the recipient has won a large lottery prize can make them focus on how to claim the money, distracting them from realizing they never bought a lottery ticket in the first place.

P5: Commitment, Integrity, and Reciprocation - People often feel compelled to reciprocate a favour or respond to an action due to a sense of commitment to a previous interaction [11]. Additionally, they generally trust that the person they are communicating with is being honest about their needs or feelings [11]. For example, an email from someone who knows the recipient is house hunting and offers a great deal on a property that matches the recipient's criteria. To secure the property (commitment), the recipient must urgently pay a deposit (reciprocation).

They also detailed the sub-principles that these main principles encompass or intersect with. The email content analysis framework, including the coding tree and codes are based on those definitions.

## III. Identifying principles of persuasion in phishing email content

To investigate the use of principles of persuasion in phishing emails, a sample of phishing emails was selected for analysis using qualitative content analysis. The following sections outlines methodology used for the sampling process and the procedures for analysing how principles of persuasion are utilized in these emails.

### A. *Sample and criteria*

The phishing emails sample was sourced from the millersmiles online platform (millersmiles.co.uk), a widely recognized and freely accessible anti-phishing service. Known internationally, millersmiles is a prominent repository for information on spoof emails and phishing scams. It maintains an extensive archive of these scams, making it a leading resource on the internet for anti-phishing efforts [23]. Utilizing this reliable archive ensures a higher level of confidence in the authenticity and validity of the included phishing email samples.

Regarding the sampling process, a 10-year timeframe from 2014 to 2023 was chosen to mitigate biases associated with trends specific to individual years. Within this timeframe, 20 phishing emails per year were selected, a total of 200 emails were selected for analysis, with 20 emails sampled from each year. This sample size aligns with common practices [11] [14] in phishing email research, where studies typically analyse between 200 to 250 samples based on literature findings. Additionally, analysis of variability indicated that consistent patterns emerged after examining 180 samples, thus confirming the decision to include 200 samples. The emails included in the study were randomly chosen using a randomized selection method.

### B. *Analysis of email content*

The email contents underwent content analysis using NVivo 14 software. The analysis involved categorizing the phishing email samples by the year of origin. Additionally, each sample was classified based on its target: if aimed at an individual, it was labelled as Individual; if directed at an organization, it was categorized as Organization. Furthermore, the data were classified according to the goal of each phishing email sample, which included categories such as Financial Gain, Service Interruption, and Unauthorized Access. The final coding tree and code definitions were established deductively, meaning they were grounded in existing theory and based on the principles of persuasion defined by Ferreira and Teles (2019), rather than emerging inductively from the data. In the coding tree, 'Parent' codes represent broader concepts or principles that encompass a set of 'child' concepts or principles. For instance, 'Social Proof' includes 'Herd', 'Diffusion of responsibility' and ‘Moral duty’. Additionally, 'Parent' codes can also represent concepts or principles that intersect with others but are not contained within any broader concept or principle, such as 'Integrity'. 'Child' codes, therefore, refer to the specific concepts or principles that fall under 'parent' codes. In the coding process, semantic texts, specifically themes, were chosen as the units of record, meaning that content segments, including multiple excerpts from a single sentence line, could be coded into different categories. The results are presented using absolute frequencies, showing the number of references coded according to principles of persuasion, sub-principles, and attributes based on the goals of the phishing emails. Additionally, graphical representations of the data illustrate keyword frequencies through word frequency queries or word clouds. The analysis and reporting of results will emphasize absolute frequencies and percentages within categories to visually represent elements found in the sample of phishing emails.

## IV. PERSUASION IN PHISHING CONTENT

This section provides the results from the content analysis conducted on the sample of phishing email content, using the predefined deductive coding tree.

### A. *Word Frequency*

A word frequency analysis was conducted to visualize the 50 most common words or concepts in the whole phishing email content. To avoid redundancy, stemmed words (e.g. update, updated, updating) were considered in the analysis. Additionally, conceptually irrelevant words (e.g. from, what’s) were added to the stop words list. The resulting word cloud highlighted that many frequent terms suggested the use of principles of persuasion such as ‘General Deception’, with words like ‘account’ (n = 437) and ‘emails’ (n = 139) standing out. The presence of words like ‘update’ (n = 129) or ‘information’ (n = 110), implying that something is being offered to the recipient, pointed to the use of the ‘Reciprocation’ principle of persuasion. Other terms, such as ‘please,’ ‘security,’ and ‘dear,’ indicated the use of the ‘Integrity’ principle. Banking-related terms, financial institutions, well-known company names, and email service provider names were frequently employed as authority figures.

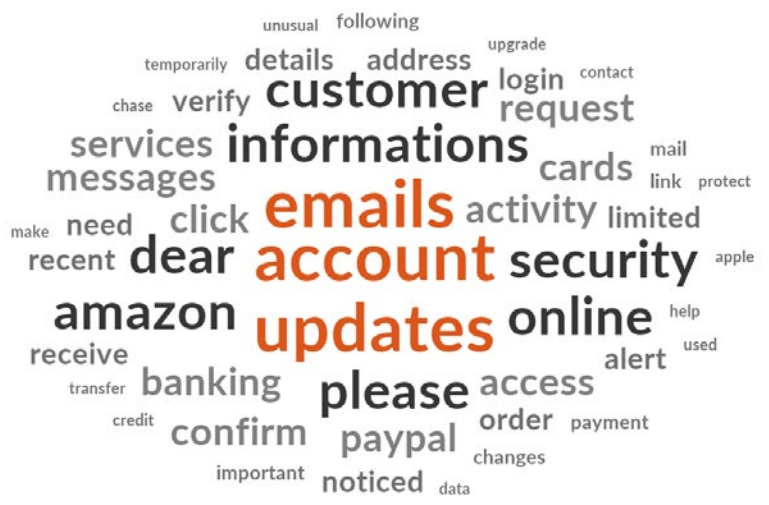

Fig. 1. Word cloud showing the most frequently occurring words in the phishing email content

### B. *The principles of persuasion in phishing email content*

The content analysis was conducted on 200 phishing email samples. By analysing the content of these emails, it was found that the most prominent principle of persuasion, based on the number of references coded, was the principle of 'Distraction,' with a total of 537 references. Those references were distributed across 171 email sample files, with some files containing more than one distraction-suggesting text element. Excerpts coded under this category indicate the use of Distraction based on Need & Greed (n=63), Overloading (n=180), Scarcity (n=103), and Strong Affect (n=191). The use of warning expressions or terms eliciting alertness (e.g., "Alert," "important," "limited") was the most frequent practice identified in this category. Additionally, some expressions were used to elicit shock and overload recipients with information to create distraction.

The second most prominent principle of persuasion identified in the analysis was 'Deception,' with 520 references across 171 files. This principle encompasses General Deception (n=228), Deceptive Relationships (n=121), and Liking & Similarity (n=171). General Deception involves manipulation without establishing relationships (e.g., "Due to a recent security issue, your account is temporarily deactivated"). Some examples of Deceptive Relationships involve creating false shared interests to foster favourability (e.g., "Dear valued customer, your account details need to be updated"). In the Liking & Similarity category, text units leverage familiarity, likability, or shared interests to persuade (e.g., "Your email service won't be affected and you'll keep all your old contacts, folders, and messages," or "Your NatWest Credit Card Online Services security details were recently changed").

The third most prominent principle of persuasion was 'Integrity,' with 507 references across 170 files. This principle includes Reciprocation (n=211), Integrity (n=106), Consistency (n=102), and Commitment (n=88). The excerpts mainly involve evoking security cues and suggesting offers of information or favours to the email recipient (e.g., "please kindly click here now to restore your old secure password," "verify apple account, this is an automated message"). These tactics aim to make recipients feel secure about replying to emails or clicking provided links, despite the potential security risks, unless there is convincing evidence to the contrary.

The fourth most prominent principle of persuasion was 'Authority,' with 196 references across 116 files. Excerpts coded under this category suggest the use of authority figures, such as known organizations, or authority cues, such as the imperative form to issue orders. References to known organizations as authority figures were particularly notable for financial or finance-related entities (e.g., PayPal, Chase, NatWest, Bank of America, FirstBank, MoneyGram, American Express). Additionally, technology companies (e.g., Apple, Microsoft, Google, Dropbox), email service providers (e.g., Yahoo, AOL, Bigpond), and other companies offering products and services (e.g., Amazon, Wells Fargo, Verizon, Netflix, FedEx, Walmart) were also used to capture the recipient's attention.

Table 1 illustrates the salience of each principle of persuasion, sorting them in descending order according to the number of references coded and the number of source files where the principles were identified. Considering the deductively defined coding tree, some principles of persuasion were identified very infrequently in the content analysis of the 200 emails, namely: Social Proof (P2); Herd (P2.1); Moral Duty (P2.3); and Diffusion of Responsibility (P2.2). Therefore, based on this analysis, conclusions cannot be drawn regarding the employment of these principles in phishing emails.

TABLE I. ABSOLUTE FREQUENCIES OF REFERENCES CODED BY PRINCIPLES.

| Main Principle | Sub-Principle | Files | References |
|---|---|---|---|
| Distraction (P4) | | 171 | 538 |
| | Strong Affect (P4.3) | 110 | 191 |
| | Overloading (P4.2) | 115 | 180 |
| | Scarcity (P4.1) | 68 | 103 |
| | Need and Greed (P4.4) | 41 | 63 |
| Deception (P3) | | 175 | 530 |
| | General Deception (P3.1) | 115 | 228 |
| | Liking and Similarity (P3.3) | 98 | 171 |
| | Deceptive Relationship (P3.2) | 101 | 121 |
| Integrity (P5) | | 170 | 508 |
| | Reciprocation (P5.4) | 129 | 211 |
| | Integrity (P5.1) | 74 | 106 |
| | Consistency (P5.2) | 67 | 102 |
| | Commitment (P5.3) | 68 | 88 |
| Authority (P1) | | 116 | 196 |
| Social Proof (P2) | | 7 | 16 |
| | Herd (P2.1) | 7 | 7 |
| | Moral Duty (P2.3) | 7 | 7 |
| | Diffusion of Responsibility (P2.2) | 1 | 1 |

Similarly, the principles of persuasion with the highest percentage of the sources' content (200 emails) were 'Deception' (27.39%), 'Distraction' (26.76%), 'Integrity' (26.60%), 'Authority' (18.15%), and 'Social Proof' (1.10%).

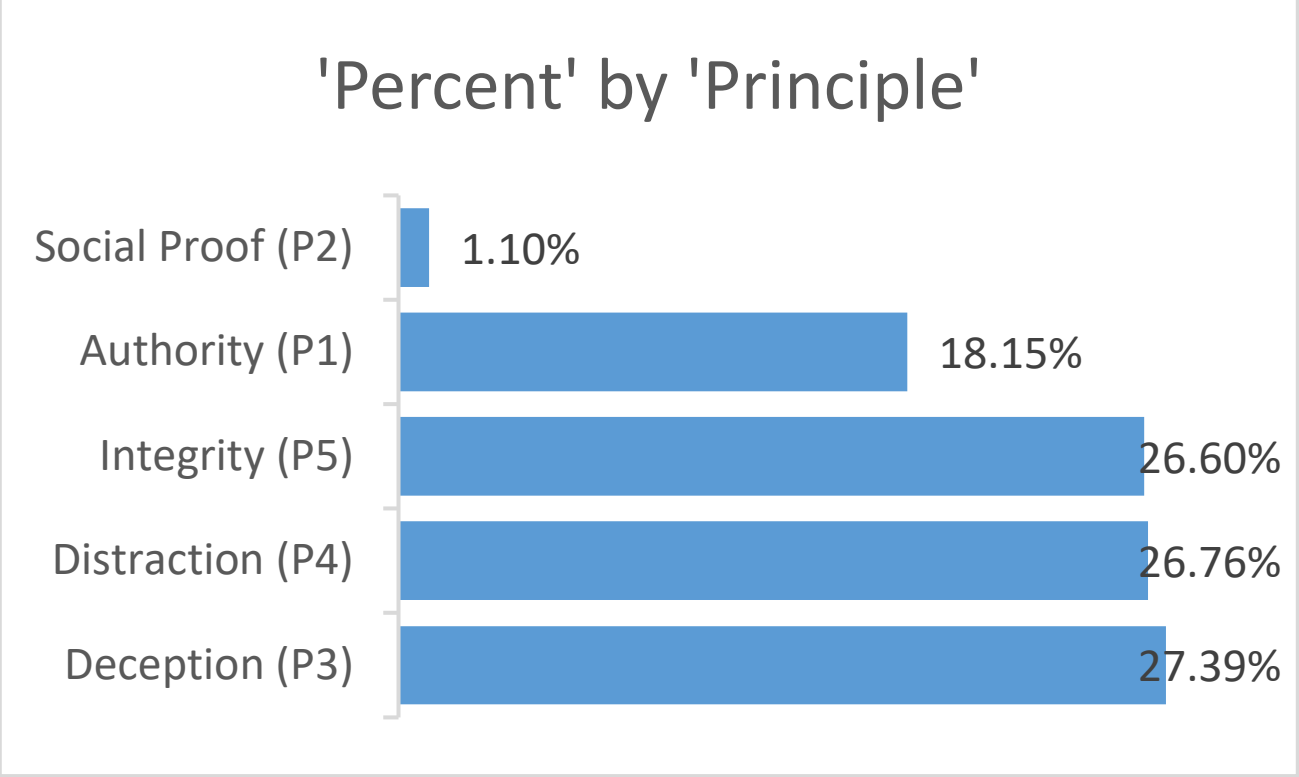

Fig. 2. Percentage of the file source (200 phishing emails) covered in each principle of persuasion category.

## V. TARGET AND GOALS OF THE PHISHING EMAILS

According to the Ontological Model of Mouton (2014) in the social engineering attack framework, a social engineering attack target can be an individual or an organization. In this study, phishing email samples were classified according to their target. All of the phishing email samples were targeted at individuals, with very few targeting organizations. Moreover, the Ontological Model of Mouton (2014) identifies three goals of social engineering attacks: Financial Gain, Unauthorized Access, and Service Disruption [24]. Among these goals, 61% of the phishing emails aimed for Unauthorized Access, and 39% aimed for Financial Gain. Very few phishing emails targeted Service Disruption, indicating that most phishing emails aim for Financial Gain and Unauthorized Access to sensitive information, which can be misused for monetary gain.

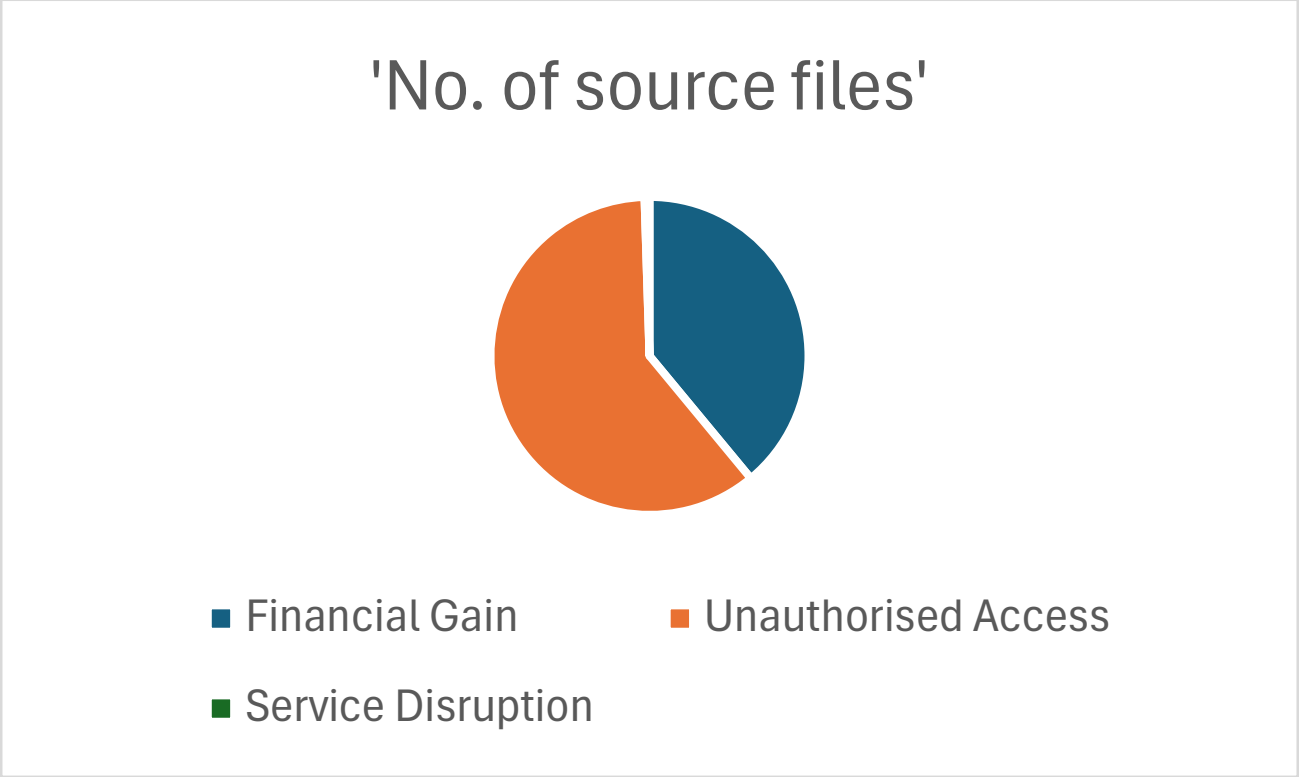

Fig. 3. Goal of the phishing email samples

## VI. Discussion

This study builds on previous research by Ferreira and Teles (2019), which synthesized three research works ( [19]; [20]; [21] to propose a unique list of principles of persuasion in social engineering. While Ferreira and Teles (2019) experimented with email subject lines, this study analysed the entire email content. Additionally, this study uniquely examined the goals and targets of phishing email samples, which is a novel contribution to the field. If this study's result is applied to the ontological model by Mouton et al. (2014), when social engineers use email as a medium for phishing, the primary targets will be individuals. The goals of these social engineering attacks are likely to be unauthorized access more often than financial gain, followed by service disruption. Furthermore, as a compliance principle, Distraction will be the most commonly used technique. Studying human persuasion in dialogues and interactions mediated by technology can aid in developing complementary methods to detect and discard phishing emails, preferably before they reach the user [11]. While luring elements in phishing emails have been previously identified, findings from empirical research in this field have often been conflicting [11]. For instance, Akbar (2014) identified 'authority' and 'scarcity' as the most common persuasion principles, whereas Ferreira et al. (2015) found 'liking and similarity' to be the most frequent. In this study, 'Distraction' was the most commonly used principle of persuasion, followed by 'Deception,' 'Integrity,' and 'Authority.' Consequently, this paper provides additional insights into the luring elements of phishing emails. Ferreira and Teles (2019) conducted a study analysing email subject lines, which was their novel approach. In the same paper, they suggested that analysing the entire email text could identify additional principles of persuasion, a gap that this paper aims to fill.

## VII. Conclusion

This study analysed the entire email content to identify principles of human persuasion within phishing emails. It demonstrates why phishing email content can be useful for automated detection and prevention. The author suggests that future solutions should focus on leveraging socio-technical principles such as distraction, deception, integrity, and authority. Future work will involve applying this methodology to other social engineering attack techniques beyond phishing emails, in conjunction with the ontological model of social engineering, to provide further insights to the research community.